\documentclass{epl}
\usepackage{ae}
\title{Charge reversal of colloidal particles}
\author{S. Pianegonda\inst{1} \and Marcia C. Barbosa\inst{1} \and Yan Levin\inst{1}}
\institute{                
  \inst{1}Instituto de F\'{\i}sica, Universidade Federal do 
Rio Grande do Sul, CP 15051, 91501-970 Porto Alegre RS, Brazil}

\begin{document}

\maketitle

\begin{abstract}
A theory is presented for the effective charge of colloidal
particles in suspensions containing multivalent
counterions.  It is shown that if  colloids are sufficiently
strongly charged, the number of condensed multivalent counterion
can exceed the bare colloidal charge leading to charge reversal.
Charge renormalization in suspensions with multivalent
counterions depends on a subtle interplay between the solvation
energies of the multivalent counterions in the bulk and near
the colloidal surface.  We find that the effective charge
is {\it not} a monotonically decreasing function of the multivalent
salt concentration.  Furthermore, contrary to the previous theories,
it is found  that except at very low concentrations, monovalent
salt hinders the charge reversal. This conclusion is in agreement
with the recent experiments and simulations.
\end{abstract}

\section{Introduction}

When a colloidal particle is placed inside a suspension containing
multivalent ions its electrophoretic mobility can 
become reversed~\cite{ToVa80,LoSaHe82}.
If this happens, an  applied electric field will produce a
drift of a colloid in the direction opposite to the one
expected based purely on its chemical 
charge~\cite{QuGo03,MaQuGa03,GaEyPa04}.  
Somehow an excessive
number of counterions must become associated 
with the colloid forming
an overcharged (charge reversed) 
complex~\cite{Le02,GrNgSh02,NgGrSh00b,BeZeHe04,GeBrPi00,SoCr01}.  
What is the cause of this curious behavior? 
 
There is a significant clues to the mechanism of charge reversal: the 
mean-field Poisson Boltzmann (PB) theory completely fails
to account for its existence~\cite{AlChGr84,MoNe00,TrBoAu02,BoTrAu02}.  
Since the PB theory does not
take into account the ionic correlations, it is reasonable
to suppose that they are the ones responsible for the colloidal
charge reversal.  Indeed, recently a number of theories 
have been advanced
to establish the mechanism through which the counterion
correlations lead to 
overcharging~\cite{QuGo03,Le02,GrNgSh02,NgGrSh00,NeJo99}.  
Unfortunately none
of the theories can fully account for the experimental findings.
While all the theories predict that addition of monovalent
salt should greatly increase the 
amount of charge reversal,  
quite opposite is found 
experimentally and in the molecular dynamics
simulations~\cite{MaQuGa03,TaGr01,Ta03}. In fact it is 
observed experimentally
that
while small concentrations of
1:1 electrolyte have little effect on the charge reversal,
larger concentrations destroy it completely~\cite{MaQuGa03}. 
Similar behavior has
also been seen in recent molecular dynamics simulations~\cite{TaGr01,Ta03}.
In this paper we will present a theory of charge reversal which
accounts for the behavior observed in the experiments and simulations.

\section{The Model}

Consider a colloidal particle of radius $a$ and charge $-Zq$, 
distributed uniformly over its surface, 
inside a suspension containing monovalent
salt at concentration $C$ and $\alpha$-valent salt at
concentration $C_{\alpha}$. All ions are
modeled as  hard spheres of diameter $a_{c}$. 
We shall assume that
both salts are strong electrolytes so that in aqueous solution
there will be  $\alpha$-valent (cation) 
counterions at concentration $C_\alpha$,
monovalent (cation) counterions at 
concentration $C$ and coions (anions) at concentration
$C+\alpha C_\alpha$.  For simplicity we will assume that all the
coions are identical.  The solvent will be treated as a
uniform continuum of dielectric constant $\epsilon$. 
A strong electrostatic interaction between the colloid and the
counterions will result in their mutual association.  If the
counterions are sufficiently strongly bound to the colloidal particle,
a new entity  --- the  colloid-counterion complex --- will be 
formed.   It is reasonable to suppose that for small
electric fields the zeta potential at the colloidal shear plane will be 
proportional to the net charge of the complex.  The Smoluchowski equation
can then be used to find the  electrophoretic mobility~\cite{FeFeNi00}. 
The goal of the present theory is then  
to calculate the number of condensed/associated counterions.

We will define the counterions as free (not-associated) if 
they are farther than distance
$\delta$ from the colloidal surface.
An ``agglomerate'' is then defined as a polyion with a
$\delta$-sheath of $n$ surrounding counterions.  
For concreteness we shall take $\delta=2$ \AA\, which 
corresponds to the characteristic  hydration radius of an ion.

Since the electrostatic 
attraction between the $\alpha$-valent counterions 
and the colloid is much
stronger than its interaction with the monovalent 
counterions and coions, it is the  multivalent
ions which are primarily responsible for the
colloidal charge renormalization. 
The size of an agglomerate is then 
determined from the minimum of the
grand potential function,
\begin{equation}
\Omega(n)=F(n)-n \mu_{0}
\label{1}
\end{equation}
where $F(n)$ is the Helmholtz free energy of the
agglomerate and $\mu_{0}$ is the bulk chemical 
potential of free  $\alpha$-ions defined later.   
The theory must provide  the expressions
for the Helmholtz free energy of the agglomerate 
and  the bulk chemical potential
of the free $\alpha$-ions.  
Let us begin with the chemical potential.

Statistical mechanics of asymmetric electrolytes still posses
an outstanding challenge to physical chemistry~\cite{AqBaFi04}. 
It is possible, however,
to gain a significant insight into the problem by appealing to the
theories advanced by Debye, Hückel and Bjerrum more than
80 years ago.  The fundamental insight of Debye and Hückel (DH) 
was that although the ions of electrolyte are {\it on average} uniformly
distributed throughout the volume of solution, there are
exist strong positional 
correlations between the ions of opposite sign~\cite{DeHu23}.
Debye and Hückel suggested that these correlations can be studied
using a linearized Poisson-Boltzmann equation (PB) --- Poisson equation
in which the ionic charge density is 
given by the Boltzmann distribution.  
Bjerrum, however,
noted that when  oppositely charged ions come into a close proximity
forming dipolar pairs,
linearization of the Boltzmann factor is no longer valid, and the DH
theory fails~\cite{Bj26}. Although for $1:1$ electrolytes 
in water at room temperature the Bjerrum
dipolar formation is only marginally relevant, for multivalent ions 
it is the primary mechanism 
responsible for the failure of the linear DH theory. 
The non-linear configurations
can be  reintroduced into the DH theory as new species --- 
dipoles and higher order clusters containing an 
$\alpha$-ion and $i=0,...\alpha$ associated anions ---
the concentrations of which, $c_i$, 
is governed by the law of mass action 
\begin{equation}
\mu_{i}=\mu_{0}+i\mu_{-}\;,
\label{2}
\end{equation}
where $\mu_{0}$ is the chemical potential of free unassociated 
$\alpha$-ions and  $\mu_{-}$ is the chemical potential of anions. 
Particle conservation imposes constraints $C_\alpha=\sum c_i$ and
$\alpha\, C_\alpha +C=c_- + \sum i c_i$, where $c_-$ is the number of
free anions.

The chemical potential of a cluster containing an $\alpha$-ion and 
$i$ associated anion is 
\begin{equation}
\beta\mu_{i}=\ln\left(\frac{c_{i}\Lambda^{3(i+1)}}{\xi_{i}}\right)+
\beta\mu_{i}^{ex}.
\label{3}
\end{equation}
The first term of Eq.~(\ref{3}) is the entropic contribution
arising from the center of mass and internal motion of the clusters. 
$\Lambda=(h/\bar m k_B T)^{1/2}$ is the de Broglie
thermal wavelength, where $\bar m$ is the geometric mean mass of the cluster,
and $\xi_{i}$ is the cluster internal partition function~\cite{LeFi96}.
For free $\alpha$-ions ($0$-clusters) $\xi_{0}=1$. 
The second
term of Eq.~(\ref{3}) is the excess chemical potential
resulting from the electrostatic interaction
between the cluster and other ionic species.
At the level of the
Debye-H\"uckel-Bjerrum (DHBj) approximation~\cite{LeFi96}, only the
interactions between charged entities contribute to the excess chemical
potential.  Thus, a neutral cluster
will not have any excess  chemical potential. 

With the non-linearities taken into account through the process
of cluster formation, the rest of the electrostatic interactions can be
treated using the linearized PB equation
\begin{equation}
\nabla^{2}\phi=\kappa^{2}\phi\;,
\label{4} 
\end{equation}
where $\kappa=\sqrt{8\pi\lambda_{B}I}$ is 
the inverse Debye length,  
$I=\frac{1}{2}[C+c_{-}+\sum_{i=0}^{\alpha}(\alpha-i)^{2}c_{i}]$ is
the ionic strength and  $\lambda_{B}=\beta q^{2}/\epsilon$ is 
the Bjerrum length.
For example, the excess chemical potential of an anion can be obtained
by integrating the Helmholtz Eq.~(\ref{4}) followed by the 
Güntelberg charging process~\cite{Gu26}
\begin{equation}
\beta\mu_{-}^{ex}=-\frac{\lambda_{B}\kappa}{2(1+\kappa a_c)} \;.
\label{5}
\end{equation}
Similarly the excess chemical potential of an $i$-cluster 
is found to be
\begin{equation}
\beta\mu_{i}^{ex}=-\frac{(\alpha-i)^{2}\lambda_{B}\kappa}{2(1+\kappa R_{i})}
\label{5a}
\end{equation}
where $R_{i}$ is the effective radius of the cluster determined from
its effective excluded volume,  
$R_0=a_c$, $R_{1}=1.191a_{c}$, $R_{2}=1.334a_{c}$, {\it etc}~\cite{LeFi96}. 

The internal partition function of an $i$-cluster is
\begin{equation}
\xi_{i}=\frac{1}{i!}\int dr_{1}^{3}...dr_{i}^{3}e^{-\beta U} \;,
\label{6}
\end{equation}
where $U$ is the  Coulomb potential.  The integral is cutoff at short
distance by the hardcore of the ions and at large distance by some
characteristic size at which the associated ions can be 
considered to belong to the same cluster.  In the  strong coupling limit, 
$\alpha \lambda_B/a_c >>1$, the precise value of the upper cutoff 
is irrelevant~\cite{LeFi96}, and the internal 
partition function can be evaluated explicitly. We find 
\begin{equation}
\xi_{1}={a_{c}}^{4}e^{\alpha\lambda_{B}/a_{c}}\frac{4\pi}{\alpha\lambda_{B}}
\label{7}
\end{equation}
\begin{equation}
\xi_{2}={a_{c}}^{9}e^{(4\alpha-1)\lambda_{B}/2a_{c}}\frac{1024\pi^2}{(4\alpha-1)^{2}\lambda_{B}^3}
\label{8}
\end{equation}
\begin{equation}
\xi_{3}={a_{c}}^{27/2}e^{(3\alpha-\sqrt 3)\lambda_{B}/a_{c}}\frac{2^{9/4}\pi^{9/2}}{{3}^{3/2}(\sqrt{2}\alpha-1)^{3}\lambda_{B}^{9/2}}
\label{9}
\end{equation}

Although it is, in principle, possible to calculate the internal partition
function of higher order clusters as well, the calculations become 
progressively more complex.  Since in this paper we are interested only 
in the case $\alpha=3$, the three internal partition functions 
given by Eqs.~(\ref{7})-(\ref{9}) are sufficient for our purpose.  

Substituting the expressions for $\mu_{-}$, $\mu_{0}$ 
and $\mu_{i}$ into the law of mass action yields
\begin{equation}
c_{i}=\xi_{i}c_{0}c_{-}^{i}e^{-\beta\mu_{i}^{ex}+\beta\mu_{0}^{ex}+
i\beta\mu_{-}^{ex}}.
\label{10}
\end{equation}
Eq.~(10) is a set  of $\alpha$ coupled algebraic equations 
which must be solved numerically to determine the distribution 
$\{c_{i}\}$ and the number of free coions $c_-$.  With these in hand,
the bulk chemical potential of free $\alpha$-ions $\mu_0$, needed for the
minimization of the grand potential can be calculated 
using Eq.~(\ref{3}). 
Since 
the electrostatic interactions are the strongest 
between the colloid and the free $\alpha$-ions, 
we have assumed that it is their condensation which
is primarily responsible for the 
colloidal charge renormalization.

Our next step is to calculate the 
free energy of an agglomerate containing a colloid and $n$ 
condensed $\alpha$-ions. The Helmholtz free energy can be written
as a sum of three terms, 
$F_n= E_n+F_n^{solv}+ F^{ent}_{n}$.  
$E_n$ is  the electrostatic free 
energy of an isolated agglomerate, $F_n^{solv}$ is the solvation free
energy that the agglomerate gains from being placed 
inside the suspension, and $F^{ent}_{n}$ is the entropic free energy
of the condensed counterions. 

The energy of an isolated agglomerate is  
\begin{equation}
\beta E_{n}=\frac{Z^2 q^2 \lambda_B}{2 a}-\frac{Z \alpha q^2 \lambda_B}{ a}+ \beta F_n^{\alpha \alpha}
\label{11a}
\end{equation}
The first term of Eq.~(\ref{11a}) is the electrostatic self energy of the
colloidal particle, the second term is the interaction 
energy between the colloid and $n$ condensed counterions, and the last
term is the electrostatic energy of interaction between the
condensed $\alpha$-ions. We can relate $F_n^{\alpha \alpha}$ to
the free energy of a spherical one component plasma (SOCP), defined as a
plasma of $n$ $\alpha$-ions moving on the surface of a sphere with a 
{\it uniform
neutralizing background},
\begin{equation}
\beta F^{SOCP}=\beta F_n^{\alpha \alpha}+ \frac{ n^2 \alpha^2 q^2 \lambda_B}{2 a}-\frac{n^2 \alpha^2 q^2 \lambda_B}{ a} 
\label{11b}
\end{equation}
In the strong 
coupling limit~\cite{Sh99a,MeHoKr01,Le02}
the free energy of the $SOCP$ is very well approximated by $\beta F^{SOCP} \approx \alpha^{2}\lambda_{B}Mn^{3/2}/2 a$, where $M=1.104$ 
is the Madelung constant. Substituting Eq.~(\ref{11b}) into
the Eq.~(\ref{11a}) the energy of an isolated agglomerate 
becomes~\cite{Sh99}
\begin{equation}
\beta E_{n}=\frac{(Z-\alpha n)^{2}\lambda_{B}}{2a}-
\frac{\alpha^{2}\lambda_{B}Mn^{3/2}}{2a}\;.
\label{11}
\end{equation}

When the agglomerate
is placed inside the electrolyte solution it gains an additional
solvation free energy which, once again, 
can be obtained using the Debye-Hückel
theory~\cite{Le02},
\begin{equation}
\beta F^{solv}_n=-\frac{(Z-\alpha n)^{2}\lambda_{B}\kappa a}{2a(1+\kappa a)}\;.
\label{12}
\end{equation}
Finally, the entropic free energy of ions inside the agglomerate is
$\beta F^{ent}_{n}=n \ln(\rho_{n}\Lambda^{3})-n$,
where $\rho_{n}=n/ 4\pi a^{2}\delta$ is the concentration 
of multivalent counterions inside the $\delta$-sheath.

In equilibrium, the number of $\alpha$-ions, $n^{*}$, inside the 
agglomerate is determined from the minimization 
of the grand potential function, 
Eq.~(\ref{1}), given by $\delta \Omega_Z=0$. The minimization is
performed at fixed $Z$. 
It is important, however, to keep in mind that not all of the
$\alpha$-ions inside the 
agglomerate are really associated with the polyion.
The way the theory is constructed, the 
region near the colloidal surface
is treated separately from the rest of electrolyte.
This leads to an artificial 
excess of the multivalent ions inside the 
$\delta$-sheath. 
The number of truly condensed counterions ($\alpha$-ion
which are inside the $\delta$-sheath precisely due to
their electrostatic coupling with the colloid)
is $n^{*}-n^*_{0}$, where the
excess $n^*_{0}$ can be found by minimizing  
the  grand potential function at $Z=0$,
$\delta \Omega|_{Z=0}=0$. 
The effective charge of the polyion-$\alpha$-ion 
complex (in units of $-q$)
is then $Z_{eff}=Z-\alpha n^{*}+\alpha n^*_{0}$.

\section{Results and Conclusions}

\begin{figure}
\centering
\resizebox{7cm}{!}{\includegraphics[clip]{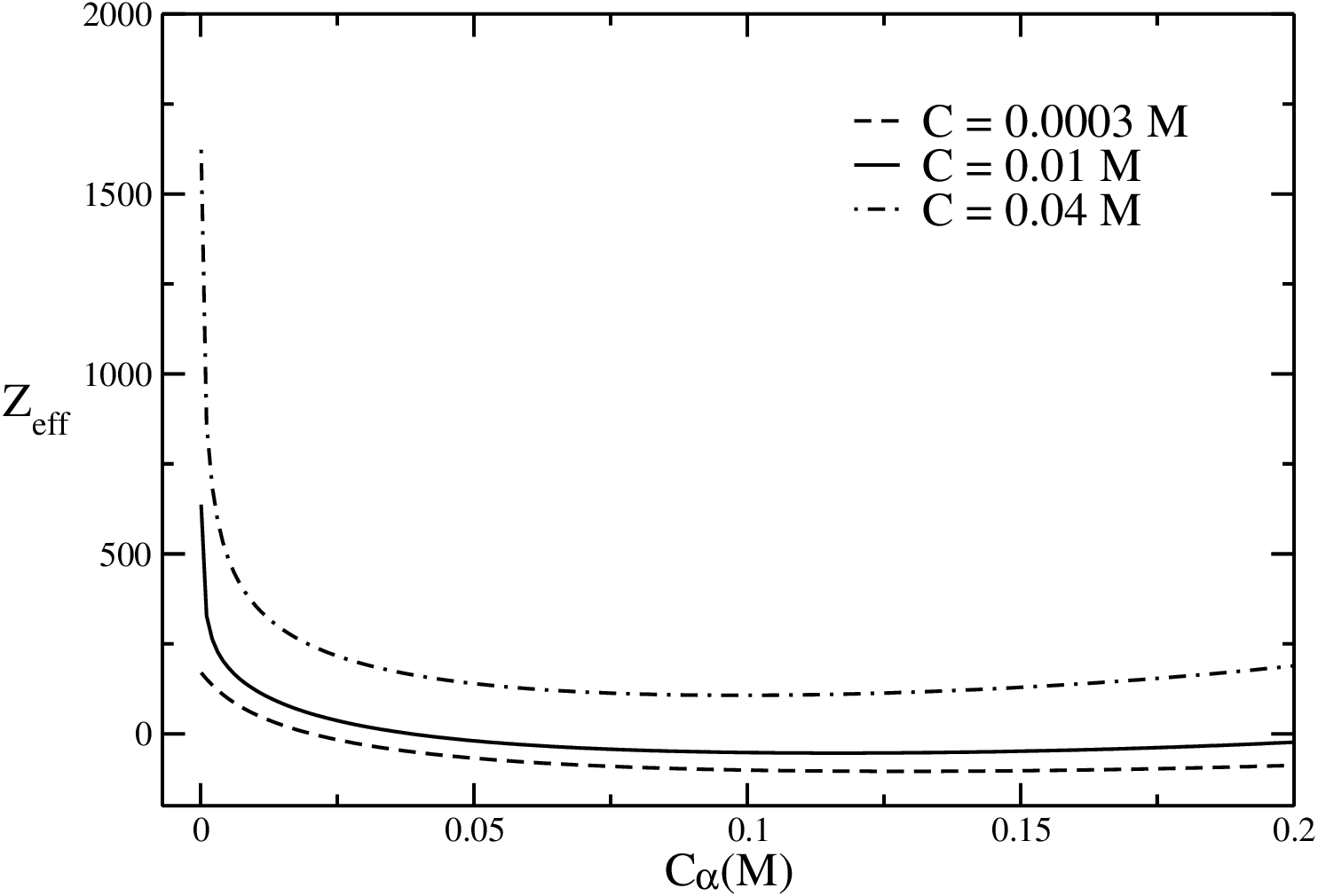}}
\caption{The effective charge as a function of 
concentration of trivalent ions ($\alpha=3$)
for a suspension containing colloidal particles 
with $Z=4000$ and $a=300$ \AA. The curves correspond
to different values of monovalent salt 
concentration with $\lambda_{B}=7.2$, $\delta=2$ \AA, $a_{c}=4$ \AA.}
\label{zeffxrhotcrhom}
\end{figure}

In figure \ref{zeffxrhotcrhom}, we present the effective 
colloidal charge as a function of concentration of
trivalent counterions ($\alpha=3$) for a suspension 
containing particles with $Z=4000$ and $a=300$\AA\, and various
concentrations of monovalent salt. 
It should be noticed that 
the charge reversal occurs only for sufficiently small 
concentrations of monovalent salt.

\begin{figure}
\centering
\resizebox{7cm}{!}{\includegraphics[clip]{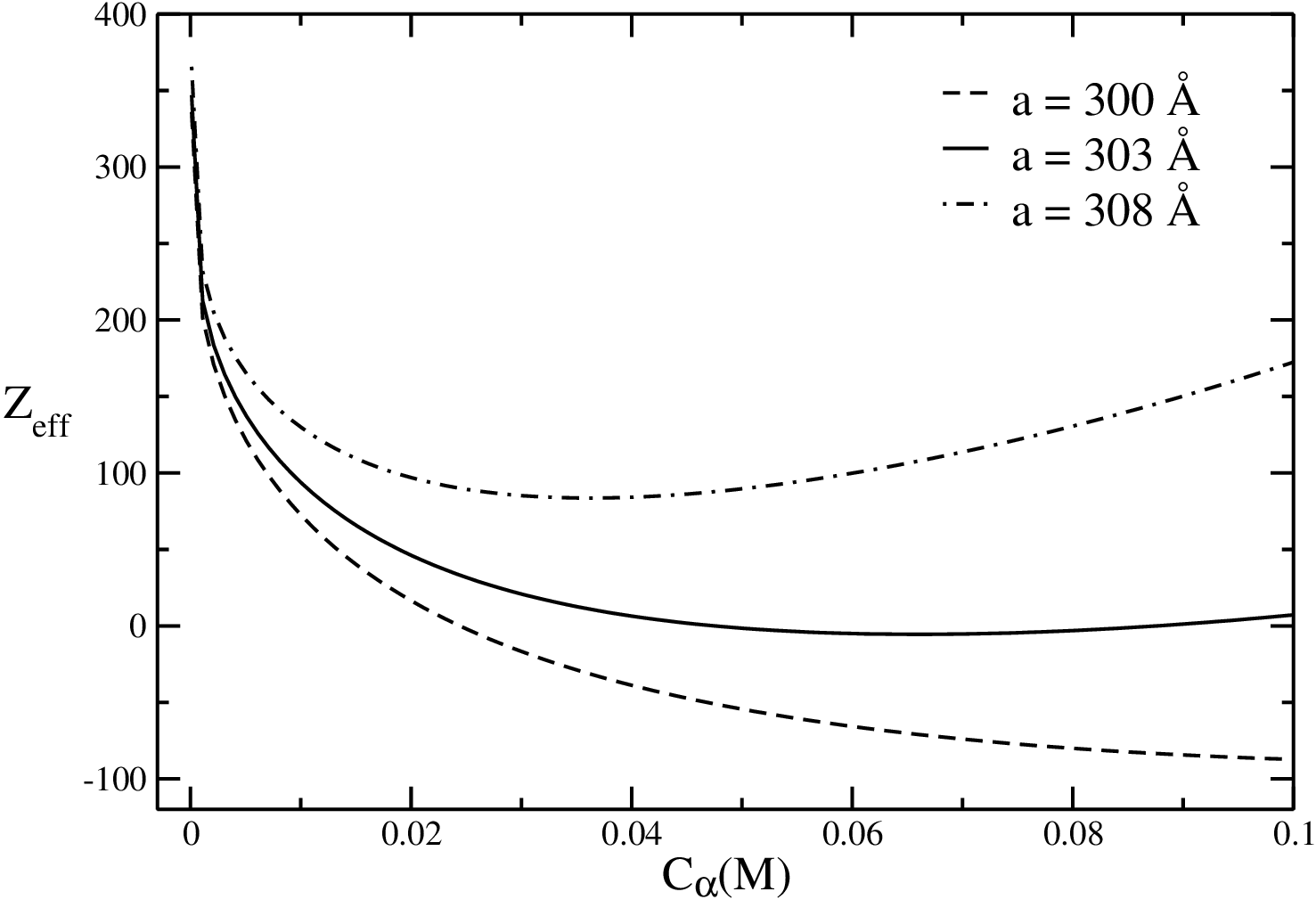}}
\caption{The effective charge as a function of 
concentration of trivalent ions ($\alpha=3$)
for a suspension of  particles with  $Z=4000$. The curves correspond
to different values of $a$ with $\lambda_{B}=7.2$, 
$\delta=2$ \AA, $C=0.003 M$ \AA, $a_{c}=4$ \AA.}
\label{zeffxrhotca}
\end{figure}
\begin{figure}
\centering
\resizebox{7cm}{!}{\includegraphics[clip]{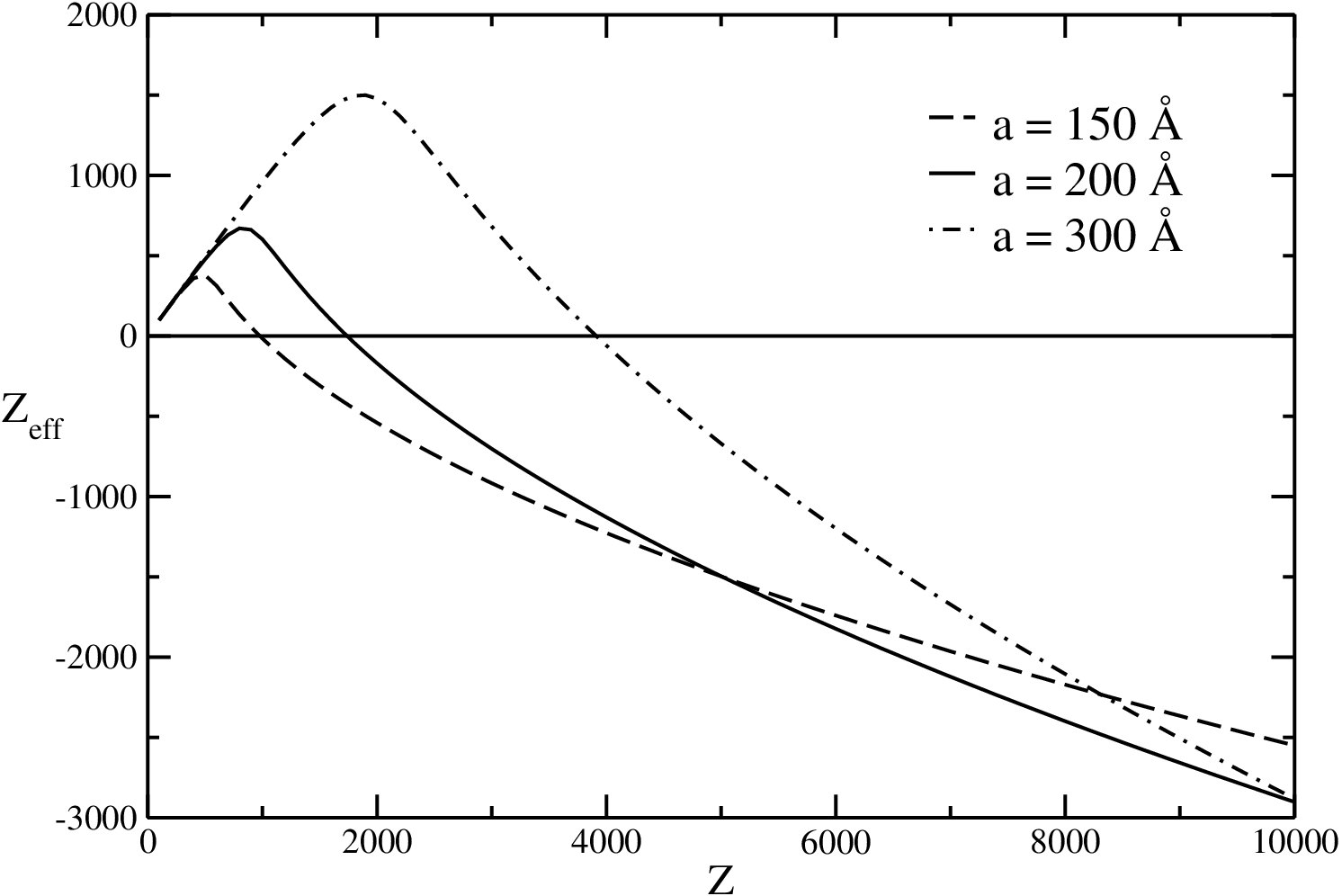}}
\caption{The effective colloidal 
charge as a function of bare charge $Z$, for 
a suspension containing monovalent salt at 
concentration $C=0.002 M$ and trivalent salt at $C_{3}=0.05 M$.
The curves correspond to different 
values of $a$ with $\lambda_{B}=7.2$, $\delta=2$ \AA\ and  $a_{c}=4$ \AA.}
\label{zeffxZcmc}
\end{figure}

\begin{figure}
\centering
\resizebox{7cm}{!}{\includegraphics[clip]{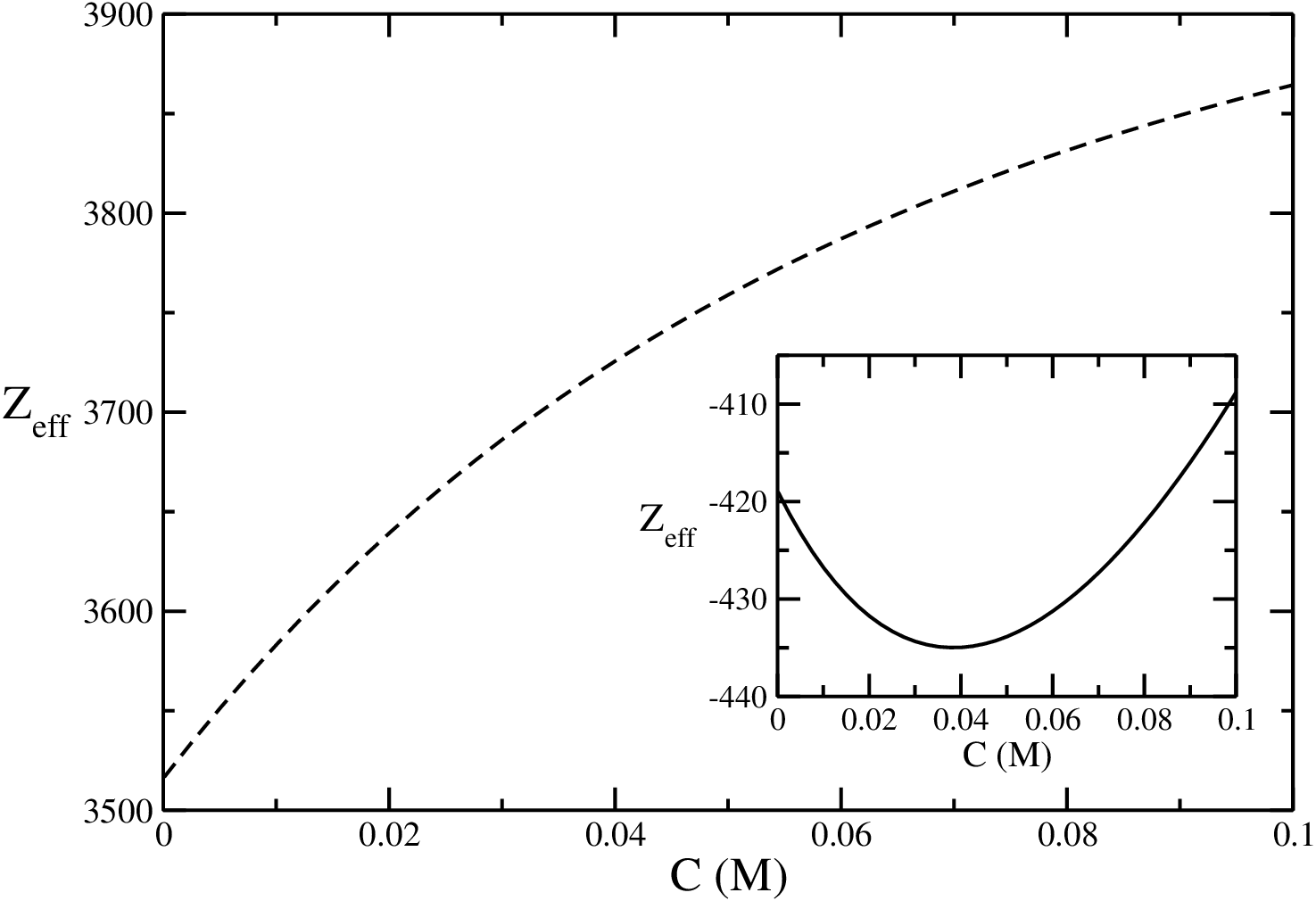}}
\caption{The effective charge of colloidal particle with $Z=4000$ and 
$a=500$ \AA\  ($\sigma<\sigma_{cr}$) at $C_{3}=0.04 M$, as a function of
concentration of monovalent salt $C$. 
The inset shows variation of $Z_{eff}$ for 
colloid with $Z=4000$ and  $a=200$ \AA\ ($\sigma>\sigma_{cr}$) 
at $C_{3}=0.01 M$, $\lambda_{B}=7.2$,
$\delta=2$ \AA, $a_{c}=4$ \AA.}
\label{zeffxmccrhot}
\end{figure}

The overcharging found in figure \ref{zeffxrhotcrhom} is a 
consequence of  strong positional correlations between the
condensed counterions. We can quantify the strength of electrostatic
correlations at the isoelectric point (when the number of condensed
counterions completely neutralizes the colloidal charge) by the plasma
parameter~\cite{Le02}
\begin{equation}
\Gamma_{iso}= \frac{\alpha^{2}q^{2}}{\epsilon dk_{B} T}
\end{equation}
where $d$ is the average separation between  $n=Z/\alpha$ 
condensed $\alpha$-ions. 
Furthermore, since 
$n \pi (d/2)^{2}=4\pi a^{2}$, the coupling strength becomes,
\begin{equation}
\Gamma_{iso}=\alpha^{3/2}\lambda_{B}\frac{\sqrt{Z}}{4a}=
\frac{1}{2}\alpha^{3/2}\lambda_{B}\sqrt{\pi \sigma}, 
\end{equation}
where $\sigma=Z/4 \pi a^2$ is the colloidal surface charge density.

The figure \ref{zeffxrhotca} shows the effective
colloidal charge as a function of concentration of
trivalent ions for colloids with $Z=4000$ 
and various sizes. As expected overcharging is 
possible only for sufficiently large 
colloidal surface charge density~\cite{Le04}.
For trivalent ions of  $a_c=4$ \AA, 
we find that the charge reversal can take place {\it if and only if }
$\Gamma_{iso} > 1.95$ or equivalently when 
the colloidal surface charge density is 
$\sigma>\sigma_{cr}\equiv 0.18/\lambda_B^2$.  
This  is a  {\it necessary} but {\it not sufficient} condition. 
For colloids with  $Z$ and  $a$ satisfying  
$\sigma>\sigma_{cr}$, the overcharging will occur only if  
the monovalent salt concentration 
is below the critical threshold $ C< C_{cr}(Z,a)$.
The critical salt concentration 
$C_{cr}(Z,a)$ is a function of {\it both} colloidal 
charge and size, and is not simply a function of $\sigma$. 
Figure \ref{zeffxZcmc} shows  
the effective colloidal charge as a function of the 
bare charge. We note that the effective charge does not 
saturate, as predicted by the PB theory~\cite{AlChGr84},
but instead reaches a maximum  and then falls
off sharply eventually going through the isoelectric point.

Finally,  the 
dependence of $Z_{eff}$ on the amount of monovalent salt 
is presented in  figure \ref{zeffxmccrhot}.
We see that for $\sigma<\sigma_{cr}$ (in the absence of overcharging),  
screening of electrostatic interactions
by 1:1 electrolyte results in a diminished
counterion-colloid association. On the other hand
if, in the absence of  1:1 electrolyte the colloid-counterion 
complex is already overcharged, addition of a small amount
of monovalent salt 
results in a slight increase of the  charge reversal, 
see the inset of Fig. \ref{zeffxmccrhot}.  
A further rise of the concentration of   
1:1 electrolyte, however, leads to a decline of the  
overcharging.  
This behavior, also observed experimentally~\cite{MaQuGa03}, 
is different from the predictions of
other theories. For example, Nguyen et al.
find that addition of large concentrations of   
monovalent electrolyte should lead to a giant charge reversal 
resulting from the screening of the 
electrostatic self energy of the overcharged  
polyion-counterion complex~\cite{NgGrSh00b}.
Contrary to this, we see that for large salt concentrations
the multivalent ions  prefer to 
be solvated in the bulk electrolyte instead of the colloidal surface.
In the bulk they gain favorable correlational energy from the
interactions with the oppositely charged coions which are 
depleted from the 
colloidal surface~\cite{TaGr01,Ta03}.  
Thus, in order for a theory to
consistently predict the conditions of 
colloidal overcharging
it must first account well for the thermodynamics of 
the bulk electrolyte solution.

\acknowledgments
We acknowledge the financial support from 
Brazilian agencies CNPq and Capes.

\bibliographystyle{prsty}
\bibliography{references}

\end{document}